\documentclass[apj,numberedappendix]{emulateapj}
\usepackage{natbib,apjfonts}

\begin{document}

\normalsize

\title{Neutron Star Crustal Interface Waves}

\author{Anthony L. Piro}
\affil{Department of Physics, Broida Hall, University of California
	\\ Santa Barbara, CA 93106; piro@physics.ucsb.edu}

\and

\author{Lars Bildsten}
\affil{Kavli Institute for Theoretical Physics and Department of Physics,
Kohn Hall, University of California
	\\ Santa Barbara, CA 93106; bildsten@kitp.ucsb.edu}

\begin{abstract}

  The eigenfrequencies of nonradial oscillations are a powerful
probe of a star's interior structure. This is especially true
when there exist discontinuities such as at the neutron star
(NS) ocean/crust boundary, as first noted by McDermott, Van Horn
\& Hansen. The interface mode associated with this boundary has
subsequently been neglected in studies of stellar nonradial
oscillations. We revisit this mode, investigating its properties
both analytically and numerically for a simple NS
envelope model. We find that it acts like a shallow surface
ocean wave, but with a large radial displacement at the
ocean/crust boundary due to flexing of the crust with shear
modulus $\mu\ll P$, the pressure. This displacement lowers
the mode's frequency by a factor of $\sim(\mu/P)^{1/2}\sim0.1$
in comparison to a shallow surface wave frequency on a hard
surface. The interface mode may be excited on accreting or
bursting NSs and future work on nonradial oscillations
should consider this mode. Our work also implies an
additional mode on massive and/or cold white dwarfs with
crystalline cores, which may have
a frequency between the f-mode and g-modes, an
otherwise empty part of the frequency domain.

\end{abstract}

\keywords{stars: neutron --- stars: oscillations --- white dwarfs}

\section{Introduction}

  A wide variety of nonradial oscillations have been previously
investigated for neutron stars (NSs) with the majority of these
studies focusing on surface g-modes (oscillations whose
restoring force is gravity; McDermott, Van Horn \& Scholl 1983;
Finn 1987;
McDermott \& Taam 1987; McDermott, Van Horn \& Hansen 1988,
hereafter MVH88; McDermott 1990; Strohmayer 1993;
Bildsten \& Cutler 1995, hereafter BC95;
Bildsten, Ushomirsky \& Cutler 1996;
Strohmayer \& Lee 1996; Bildsten \& Cumming 1998;
Piro \& Bildsten 2004) because their frequencies are similar to
many of the oscillations observed from NSs (for example burst
oscillations; Muno et al. 2001). These modes are confined to
outer regions of the NS surface due to the shear modulus of the
NS crust (as illustrated by BC95 and further discussed here), so
that their frequencies are independent of the deep NS crust
composition and small amounts of energy will lead to
large amplitudes. However one mode has consistently been
neglected, which is the wave analogous to a shallow water wave
in the outermost parts of the star. Unlike a global surface wave
(i.e. f-mode), we expect this wave to be confined to the NS ocean
just like the g-modes, but a correct prediction of its frequency
requires proper treatment of its eigenfunction at the ocean/crust
boundary.

  A study of nonradial oscillations on non-rotating, non-magnetic
NSs was presented by MVH88 for three-component NS models
(consisting of an ocean, crust, and core), examining g-modes,
acoustic oscillations (p-modes), and f-modes. In addition, MVH88
reported a new set of oscillations, which they called
``interface modes.'' These modes are unique in that they ``live''
(have amplitudes concentrated) around discontinuities in the NS
structure. For their three-component model, they found these modes
at both the ocean/crust interface and the crust/core interface. We
now undertake a detailed investigation of the ocean/crust interface
mode so that it may be used for accreting NSs and in the other
astrophysical context where a shear modulus is important,
crystalline white dwarfs (WDs) (e.g. Hansen \& Van Horn 1979,
hereafter HV79; Montgomery \& Winget 1999).

  To focus our analysis we consider the interface mode for a
two-component NS model consisting of an ocean and crust.
We use this model because of the simplification provided by the
plane parallel geometry of the outer NS layers. The most important
property of the interface mode is how it couples with the top of
the NS crust (Figure 1). MVH88 showed that the interface mode has a
large and approximately constant transverse velocity in the ocean,
similar to a shallow water wave. Indeed this mode can be thought of
as a surface wave riding on a compressible crust. It is therefore
not a toroidal mode as one might expect if the interface wave were
like a shear wave (see Appendix A). As its
pressure perturbation rolls over the crust's surface, the crust is
compressed and sinks. This flexing causes the interface mode to
have a large, negative radial displacement at the discontinuity
(see the right diagram of Figure 1) in contrast with the typical
boundary condition for surface water waves, which treats the crustal
boundary as solid with no radial displacement
(see the left diagram of Figure 1). Since the pressure
perturbation also corresponds to a small, positive swell at the top
of the ocean, the radial displacement eigenfunction must necessarily
have a single node within the ocean.
\begin{figure*}
\epsscale{0.96}
\plotone{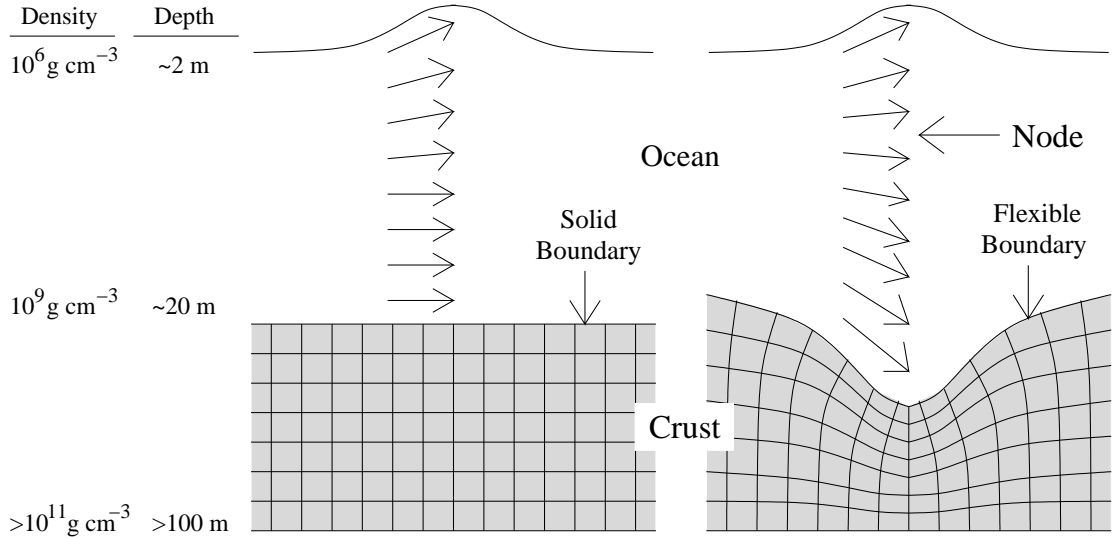}
\figcaption{A schematic diagram of an ocean surface wave showing
how it is altered by the flexing of the crust. The vectors show
the wave velocity. In both cases the
transverse velocity is approximately constant within the ocean and
dies off quickly into the crust. If the crust is treated as if it is ``solid''
the radial displacement goes to zero at the ocean floor (left diagram).
As shown on the right, the interface mode has a large, negative radial
displacement at the crust which is actually larger than the corresponding swell
near the ocean surface. This implies that its radial eigenfunction must
necessarily have a single node within the ocean.}
\end{figure*}

  This simple picture describes
the interface mode quite well. The frequency depends on the properties
at the base of the ocean, like a surface ocean wave, but with a
frequency that is altered by the non-zero radial displacement at
the ocean/crust boundary. This boundary condition
reduces the frequency by $\sim(\mu/P)^{1/2}\sim0.1$ from what is
expected if the crust were ``solid,'' where $\mu$ is the shear modulus
and $P$ is the pressure at the top of the crust. Most of the energy
of the mode is in the ocean, just as in a shallow surface wave.

  We derive the adiabatic perturbation equations for a plane parallel
layer in hydrostatic balance with a shear stress in \S 2. These are
used in \S 3 to analytically estimate the interface mode's radial
eigenfunction and frequency.  We follow this with a numerical study
in \S 4 that confirms our analytic results and also provides
more detail about the eigenfunctions.
% We then extend
% this analysis to a slightly more complicated NS envelope configuration
% to see how the interface mode changes when there exist g-modes
% of comparable frequency. In this case we find avoided crossings
% between the interface mode and other g-modes, but that our analytic
% predictions still hold.
We conclude in \S 5 with a discussion of possible uses
for the interface mode in astrophysical contexts. Maybe most exciting
among these is an additional mode of oscillation in WDs
with crystalline cores.

\section{Adiabatic Perturbation Equations}

  Since the pressure scale height in the NS ocean,
$h=P/(\rho g)\approx (2-10)\times10^2\textrm{ cm}$, is always much
less the NS radius, $R\approx 10^6\textrm{ cm}$, we approximate
the surface as having constant gravitational acceleration, $g=GM/R^2$
(neglecting general relativity), and plane parallel geometry. We use $z$
as our radial coordinate and $x$ as the transverse coordinate. An
additional useful coordinate in this geometry is the column depth, denoted
as $y$ (defined by $dy=-\rho dz$ and with cgs units of
$\textrm{ g cm}^{-2}$).

  The interface mode's period is set by physics at the ocean/crust boundary,
and is insensitive to the outer boundary condition. At these depths the thermal
timescale is longer than the mode period (since $t_{\rm th}\approx c_pyT/F$
and $y$ is large; where $c_p$ is the specific heat capacity and $F$ is the
flux) so that the perturbations are adiabatic. The governing equations of the
envelope are conservation of mass,
$\partial \rho/\partial t+\mbox{\boldmath$\nabla$}\cdot ( \rho {\bf v} ) = 0$,
and momentum,
\begin{eqnarray}
	\rho \left( \frac{\partial}{\partial t}
		+ v_j \frac{\partial}{\partial x_j} \right) v_i
	= \frac{\partial}{\partial x_j} \sigma_{ij}
		- \rho g \hat{\bf z},
\end{eqnarray}
where $\sigma_{ij}$ is the stress tensor, and $\hat{\bf z}$ is a radial unit
vector (repeated indices imply the Einstein summation convention unless
noted otherwise). We neglect the NS spin so that we can focus on the effects
of the crust. Rotation can be simply included using the
``traditional approximation'' (e.g. Bildsten, Ushomirsky \& Cutler 1996).

  We make Eulerian perturbations of the conservation equations,
substituting for the dependent variables $Q \rightarrow Q_0 + \delta Q$,
where $Q_0$ denotes the static background (this subscript is dropped in
subsequent expressions).  Perturbations are assumed to
have the form
$\delta Q= \delta Q(z)\exp(ik_x x+ i\omega t)$, where $k_x^2=l(l+1)/R^2$
is the transverse wavenumber and $\omega$ is the mode frequency.
These Eulerian perturbations are related to Lagrangian perturbations
by $\Delta Q = \delta Q + \mbox{\boldmath$\xi$} \cdot \mbox{\boldmath$\nabla$}Q$.
Since the background model has no fluid motion we substitute
${\bf v} \rightarrow \delta {\bf v} = d\mbox{\boldmath$\xi$}/dt=i\omega\mbox{\boldmath$\xi$}$,
where $\mbox{\boldmath$\xi$}$ is the local Lagrangian displacement. We
use the Cowling approximation and neglect perturbations of the gravitational
potential (Cowling 1941), an excellent approximation in the thin outer layers
of the NS. Keeping terms
of linear order, we find
\begin{eqnarray}
	\delta \rho + \mbox{\boldmath$\nabla$} \cdot
		( \rho \mbox{\boldmath$\xi$} ) = 0,
\end{eqnarray}
for continuity.

  In making linear perturbations of equation (1) we include the effects
of a finite shear modulus (such as in the NS crust). Relative displacements
then introduce a shear component to the Lagrangian stress tensor given by
(Landau \& Lifshitz 1970).
\begin{eqnarray}
	\Delta\sigma^s_{ij}
	= \mu\left(\frac{\partial\xi_i}{\partial x_j}
		+ \frac{\partial\xi_j}{\partial x_i}
		- \frac{2}{3} (\mbox{\boldmath$\nabla$}
		\cdot\mbox{\boldmath$\xi$})\delta_{ij}\right),
\end{eqnarray}
where $\mu$ is the crystalline shear modulus as
calculated by Strohmayer et al. (1991) for a
classical one-component plasma (OCP),
\begin{eqnarray}
	\mu = \frac{0.1194}{1+0.595(173/\Gamma)^2}
		\frac{n_i(Ze)^2}{a},
\end{eqnarray}
where $n_i$ is the ion number density, $a=(3/4\pi n_i)^{1/3}$
is the average ion spacing, $Z$ is the charge per ion, and
\begin{eqnarray}
	\Gamma \equiv \frac{(Ze)^2}{ak_{\rm B}T}
	= \frac{127}{T_8/4}\left(\frac{Z}{30}\right)^2
	\left( \frac{64}{A}\right)^{1/3}
	\left(\frac{\rho}{10^9\textrm{ g cm}^{-3}} \right)^{1/3},
\end{eqnarray}
is the dimensionless parameter that determines the liquid/solid transition
for the OCP, where $k_{\rm B}$ is
Boltzmann's constant, $T_8\equiv T/10^8\textrm{ K}$, and $A$ is the
number of nucleons per ion. This transition occurs
at $\Gamma\approx173$ (Farouki \& Hamaguchi 1993 and references
therein, following the work of Brush, Sahkin \& Teller 1966). Since
the pressure in the crust is dominated by degenerate electrons we
rewrite $\mu$ as
\begin{eqnarray}
	\frac{\mu}{P}
	= \frac{1.4\times10^{-2}}{1+0.595(173/\Gamma)^2}
	\left( \frac{Z}{30} \right)^{2/3}.
\end{eqnarray}
In the crust $\mu/P$ is fairly independent of temperature
(except for a small
dependence due to the factor of $\Gamma$ in the denominator), so we
typically substitute $\mu_0\equiv\mu/P$ and assume that $\mu_0$
is constant with depth.

  In the adiabatic limit, $\Delta P/P = \Delta \rho/(\Gamma_1 \rho)$,
where $\Gamma_1 \equiv (\partial\ln P/\partial\ln\rho)_{s}$ is the adiabatic
exponent. The Lagrangian stress tensor is
\begin{eqnarray}
	\Delta \sigma_{ij} = P\Gamma_1
	(\mbox{\boldmath$\nabla$} \cdot\mbox{\boldmath$\xi$})\delta_{ij}
		+ \Delta\sigma^s_{ij},
\end{eqnarray}
and the Eulerian stress tensor is
\begin{eqnarray}
	\delta\sigma_{ij} = \Delta\sigma_{ij} - \rho g \xi_z\delta_{ij},
\end{eqnarray}
so that the perturbed momentum balance equation is
\begin{eqnarray}
	-\rho\omega^2\xi_i =
		\frac{\partial}{\partial x_j}\delta\sigma_{ij}
		- \delta\rho g \hat{\bf z}.
\end{eqnarray}
Equations (2) and (9) describe the nonradial oscillations.

  The fluid viscosity in the ocean barely resists shear so that
the Lagrangian stress tensor has only diagonal components,
$\Delta\sigma_{ij}=-\Delta P\delta_{ij}$. Breaking equation (9) into
radial and transverse components,
\begin{eqnarray}
	- \rho \omega^2 \xi_z
        	&=& - \frac{d \delta P}{dz} - g \delta \rho,
	\\
	- \rho \omega^2 \xi_x
                &=& - ik \delta P.
\end{eqnarray}
Combining equations (2), (10), and (11) results in the standard nonradial
oscillation equations for an inviscid fluid in hydrostatic balance and plane
parallel geometry,
\begin{eqnarray}
	\frac{d\xi_z}{dz}
        - \frac{\xi_z}{\Gamma_1h}
        = 
        \left( \frac{ghk_x^2}{\omega^2}- \frac{1}{\Gamma_1 } \right)
        \frac{\delta P}{P},
        \\
        \frac{d}{dz} \frac{\delta P}{P}
        - \left( 1 - \frac{1}{\Gamma_1}  \right) \frac{1}{h} \frac{\delta P}{P}
        = \left( \frac{\omega^2}{g} - \frac{N^2}{g} \right) \frac{\xi_z}{h},
\end{eqnarray}
where
\begin{eqnarray}
	N^2 = -g\left( \frac{d\log\rho}{dz}
			-\frac{1}{\Gamma_1}\frac{d\log P}{dz}\right),
\end{eqnarray}
is the Brunt-V\"{a}is\"{a}l\"{a} frequency which measures the internal
buoyancy of the envelope. Normal modes of oscillation are found
by assuming $\Delta P=0$ at the top boundary (a fluid element riding
at the top feels no pressure perturbation) and then shooting for the
bottom boundary condition. This top condition, though not unique,
is fairly robust since little mode energy resides in the low density
upper altitudes of the ocean (BC95).

  When $\mu\neq0$, expanding equation (9) in both the
radial and transverse directions, and using the adiabatic condition
and equation (2) to eliminate $\delta\rho$ and $\delta P$, results in
\begin{eqnarray}
	-\frac{d^2\xi_z}{dz^2} \left( \Gamma_1 + \frac{4\mu}{3P}\right)
	&=&\xi_z \left( \frac{\omega^2}{gh} - \frac{\mu k_x^2}{P}\right)
	\nonumber
	\\
	&&+ \frac{d\xi_z}{dz} \left( \frac{4}{3P}\frac{d\mu}{dz}
		-\frac{\Gamma_1}{h} + \frac{d\Gamma_1}{dz} \right)
	\nonumber
	\\
	&&+ik_x\xi_x \left[ \frac{(1-\Gamma_1)}{h}
		- \frac{2}{3P}\frac{d\mu}{dz}
		+ \frac{d\Gamma_1}{dz}\right]
	\nonumber
	\\
	&&+ ik_x\frac{d\xi_x}{dz}\left( \frac{\mu}{3P} + \Gamma_1 \right),
\end{eqnarray}
and
\begin{eqnarray}
	-\frac{\mu}{P} \frac{d^2\xi_x}{dz^2}
	&=&\xi_x \left( \frac{\omega^2}{gh} 
		- \frac{4k_x^2\mu}{3P} - \Gamma_1k_x^2\right)
		+ \frac{1}{P}\frac{d\mu}{dz}\frac{d\xi_x}{dz}
	\nonumber
	\\
	&&+ik_x\xi_z \left( \frac{1}{P}\frac{d\mu}{dz}-\frac{1}{h} \right)
		+ik_x\frac{d\xi_z}{dz} \left( \frac{\mu}{3P}+\Gamma_1 \right),
\end{eqnarray}
the crustal mode equations derived by BC95.

  The second derivatives in equations (15) and (16) imply additional
boundary conditions at the ocean/crust interface. We require that
$\xi_z$ is continuous, so that there is no ``space'' between the two
layers, and that $\Delta\sigma_{zz}$ is continuous, otherwise there is
an infinite radial acceleration across the boundary. Furthermore,
$\Delta\sigma_{xz}=0$ because the ocean, with no shear modulus,
cannot sustain finite transverse shear. We also require the
additional boundary conditions that $\xi_z\approx\xi_x\approx0$ at
some depth deep within the crust. Though this is not necessarily the
case, it is helpful for estimating the interface mode's frequencies.

\section{Analytic Estimates}

  We now use these differential equations to analytically
derive the frequencies and eigenfunctions of the interface mode.
As we explain in Appendix A, the interface wave is not well
described by WKB analysis, so we proceed with use of
analogies to shallow ocean surface waves. In
\S 3.1, we solve equations (12) and (13) for a surface wave,
and consider how its frequency changes when $\xi_z\neq0$ at the ocean
floor (as suggested by Figure 1). In \S 3.2 we solve for the radial
crustal eigenfunction, which we use in $\S 3.3$ to connect
to the ocean perturbation and find the
radial displacement at the interface. In \S 3.4 we summarize
our frequency prediction for the interface wave and
show how it depends on the temperature and composition of
the NS crust.

\subsection{Surface Wave Analytics}

  We assume that the surface wave in the ocean has a constant transverse
displacement so that $\xi_x=\xi_{x,\rm t}$ and $d\xi/dz=0$ (where we use
the subscript ``t'' to denote the ``top'' of the ocean). At the top the
boundary condition is $\Delta P = 0 = \delta P - \xi_z\rho g$, and by
combining this with transverse momentum conservation, equation (11),
we relate the surface radial and transverse displacements
\begin{eqnarray}
	\xi_{z,\rm t} = \frac{\omega^2}{ikg}\xi_{x,\rm t}.
\end{eqnarray}
Using equation (11) to eliminate $\delta P$ from equation (12), and then
eliminating $\xi_{x,\rm t}$ using equation (17),
\begin{eqnarray}
	\frac{d\xi_z}{dz}
        - \frac{\xi_z}{\Gamma_1h}
        = \frac{\xi_{z,\rm t}}{h}
        \left( \frac{ghk_x^2}{\omega^2}- \frac{1}{\Gamma_1 } \right).
\end{eqnarray}
The electrons in the deep NS ocean are degenerate and relativistic
so that the equation of state can be approximated by a $P=K\rho^{4/3}$
polytrope, where $K$ is a function of the mean molecular weight per
electron. Using hydrostatic balance, the pressure scale height is
\begin{eqnarray}
	h(z) = \frac{1}{4}(z_{\rm t} - z),
\end{eqnarray}
where $z_{\rm t} \equiv 4K\rho^{1/3}_{\rm c}/g$, the subscript ``c'' denotes
the top of the ``crust'', and $z=0$ at the ocean/crust interface (we also
make the approximation that $\rho_{\rm c}\gg\rho_{\rm t}$). Substituting
this for $h$ in equation (18) and integrating gives,
\begin{eqnarray}
	\frac{\xi_z(z)}{\xi_{z,\rm t}} 
	= 1 - \left(\frac{4}{\Gamma_1}+1\right)^{-1}
		\frac{gk_x^2}{\omega^2}\left(z_{\rm t} - z\right),
\end{eqnarray}
where the constant of integration is set to assure that $\xi_z=\xi_{z,\rm t}$
at $z = z_{\rm t}$.

  The frequency of this surface wave is set by specifying the boundary
condition $\xi_{z,\rm c}\equiv\xi_z(z=0)$. Substituting this into equation
(20) we find a frequency
\begin{eqnarray}
	\omega^2
	= \frac{(4/\Gamma_1+1)^{-1}gk_x^2 z_{\rm t}}{1-\xi_{z,\rm c}/\xi_{z,\rm t}}.
\end{eqnarray}
To understand this result it is helpful to compare it to the frequency for a
mode with $\xi_{z,\rm c}=0$ as expected for a solid floor, which we denote
as $\omega_0$. This results in
\begin{eqnarray}
	\omega^2_0 = \left(\frac{4}{\Gamma_1}+1\right)^{-1}
		gk_x^2 z_{\rm t}.
\end{eqnarray}
For $\Gamma_1=4/3$ we find that $\omega^2_0 = gh_{\rm c}k_x^2$,
where $h_{\rm c}$ is the scale height at the bottom of the ocean, just as
expected for the shallow surface wave. The interface mode's frequency
is then
\begin{eqnarray}
	\omega^2 = \omega^2_0/(1-\xi_{z,\rm c}/\xi_{z,\rm t}).
\end{eqnarray}
We expect $\xi_{z,\rm c}/\xi_{z,\rm t}<0$ because a positive pressure
perturbation will compress the crust (see the discussion in \S 1 and
Figure 1). Equation (23) shows that flexing of
the crust results in a {\it lower frequency} than what would be expected
if the ocean floor were solid.

\subsection{The Radial Crustal Eigenfunction}

  To simplify equations (15), so that it can be solved analytically,
we order the terms and drop those which are small. We first relate
the ordering of the transverse and radial displacements within the crust.
The ocean and crust slip with respect to one another so that the
transverse displacement changes discontinuously at the boundary.
We therefore set $\xi_{x,\rm c} = \lambda\xi_{x,\rm t}$ where $\lambda$
is a dimensionless constant which we call
the ``discontinuity eigenvalue.'' It determines how $\xi_x$ changes at
the interface, and is an eigenvalue set by the bottom boundary condition
within the crust. We then rewrite equation (17) as
\begin{eqnarray}
	\xi_{x,\rm c} = \frac{igk_x}{\omega^2}\lambda\xi_{z,\rm t}.
\end{eqnarray}
At the top of the crust, the displacements must satisfy $\Delta\sigma_{xz}=0$
so that
\begin{eqnarray}
	\mu_0\frac{d\xi_z}{dz}+\mu_0ik_x\xi_z=0,
\end{eqnarray}
where we have used $\mu_0\equiv\mu/P$.
Assuming that $\mathcal{O}(d\xi_x/dz)\sim\mathcal{O}(\xi_{x,\rm c}/h)$
within the crust, we combine equations (24) and (25) to find
\begin{eqnarray}
	\mathcal{O}(\lambda)\sim\mathcal{O}
		\left(\frac{\omega^2h}{g} \frac{\xi_z}{\xi_{z,\rm t}}\right).
\end{eqnarray}
The largest reasonable frequency is that of a discontinuity
g-mode at the boundary ($\omega^2\approx ghk_x^2$)
for which the terms in equation (15) with
$\omega^2$ appearing are negligible. Using order of
magnitude estimates of $k_xh\sim10^{-4}-10^{-3}$ and
$\mu_0\approx10^{-2}$, from equation (6),
we identify the two highest order terms of equation (15),
resulting in
\begin{eqnarray}
	\frac{d^2\xi_z}{dz^2} = \frac{1}{h}\frac{d\xi_z}{dz},
\end{eqnarray}
as the simplified radial differential equation.

  Since $\Gamma_1\approx4/3$, we introduce a new radial
variable, $s$, defined by $s=4h$ which is set equal to $s_{\rm c}$
at the ocean/crust interface and increases going down so that
$ds = - dz$. Integration of equation (27) results in two constants of
integration. The first is set by requiring that $\xi_z = \xi_{z,\rm c}$
at $s=s_{\rm c}$ and the second by requiring that $\xi_z=0$
at some bottom depth, which we denote as $s_{\rm b}$,
so that
\begin{eqnarray}
	\xi_z(s) = \xi_{z,\rm c}\left[ \frac{\left( s_{\rm b}/s\right)^3-1}
			{\left(s_{\rm b}/s_{\rm c}\right)^3-1}\right].
\end{eqnarray}
If we place the lower boundary condition infinitely deep within the crust,
so that $s_{\rm b}\gg s_{\rm c}$, equation (28) simplifies to
\begin{eqnarray}
	\xi_z(s) = \xi_{z,\rm c}\left( s_{\rm c}/s\right)^3,
\end{eqnarray}
identical to what BC95 estimate for the crustal eigenfunction (in their case
in the context of a g-mode extending into the crust). 

\subsection{Connecting the Modes at the Ocean/Crust Boundary}

  We solve for $\xi_{z,\rm c}$, and estimate the mode
frequency using equation (21), by demanding that
$\Delta\sigma_{zz}$ be continuous across the ocean/crust boundary
to avoid infinite radial accelerations. Expanding
this component of the stress tensor,
\begin{eqnarray}
	\frac{\Delta\sigma_{zz}}{P} = \left( \Gamma_1-\frac{2\mu_0}{3}\right)
	\left( \mbox{\boldmath$\nabla$} \cdot\mbox{\boldmath$\xi$} \right)
	+2\mu_0\frac{d\xi_z}{dz}.
\end{eqnarray}
We evaluate this condition using the eigenfunctions we have found
for both the ocean and crust. In the ocean, above the interface,
\begin{eqnarray}
	\left(\frac{\Delta\sigma_{zz}}{P}\right)_{\rm above}
		= \frac{1}{h_{\rm c}}(\xi_{z,\rm c}-\xi_{z,\rm t}).
\end{eqnarray}
In the crust we use equation (28) to evaluate $d\xi_z/dz$. The term
involving $ik_x\xi_x$ can be ignored because it is order
$(k_xh_{\rm c})^2$ less than the $d\xi_z/dz$ term
(see Appendix B), so that
below the interface,
\begin{eqnarray}
	\left(\frac{\Delta\sigma_{zz}}{P}\right)_{\rm below}
	\approx \left(\frac{3\Gamma_1/4+\mu_0}{1-(s_{\rm c}/s_{\rm b})^3}\right)
		\frac{\xi_{z,\rm c}}{h_{\rm c}},
\end{eqnarray}
where we have substituted $s_{\rm c}=4h_{\rm c}$.
Setting equations (31) and (32) equal, we solve for
the radial displacement at the crust,
\begin{eqnarray}
	\frac{\xi_{z,\rm c}}{\xi_{z,\rm t}}
	= -\left[ \frac{3\Gamma_1/4+\mu_0}{1-(s_{\rm c}/s_{\rm b})^3}
			-1\right]^{-1}
	\approx -\frac{1}{\mu_0},
\end{eqnarray}
where the last estimate is made for $\Gamma_1=4/3$
and $s_{\rm b}\gg s_{\rm c}$.

\subsection{Interface Wave Frequency Estimate}

  From our estimate for the eigenfrequency, equation (23), along with
$\xi_{z,\rm c}/\xi_{z,\rm t}\approx-1/\mu_0$, equation (33), we find the
simple result that
\begin{eqnarray}
	\omega^2\approx\mu_0\omega_0^2,
\end{eqnarray}
The frequency of the interface mode is a factor of $\mu_0^{1/2}\sim0.1$
less than a shallow surface wave in the NS ocean.

  It is an interesting exercise to use this result to predict the interface
wave's frequency and dependence on the ocean/crust boundary's
properties. At this depth, the pressure is dominated by relativistic
and degenerate electrons with a Fermi energy of
($E_{\rm F}\gtrsim m_ec^2\gg k_{\rm B}T$)
\begin{eqnarray}
	E_{\rm F}
	=4.1\textrm{ MeV}\left(\frac{2Z}{A}\right)^{1/3}
		\left(\frac{\rho}{10^9\textrm{g cm}^{-3}}\right)^{1/3}.
\end{eqnarray}
Since the pressure is $P\approx n_eE_{\rm F}/4$, the scale height at the crust is
\begin{eqnarray}
	h_{\rm c} = 2620\textrm{ cm}
		\left(\frac{2Z}{A}\right)^{4/3}
		\left(\frac{\rho}{10^9\textrm{g cm}^{-3}}\right)^{1/3}.
\end{eqnarray}
Using $\omega_0=gh_{\rm c}k_x^2$ and substituting $\Gamma$ from equation
(5) to eliminate $\rho$, we find the interface wave has a frequency
of
\begin{eqnarray}
        \frac{\omega}{2\pi}&\approx&16.5\textrm{ Hz}
		\left(\frac{\Gamma}{173}\right)^{1/2}
		\left(\frac{T_8}{4}\right)^{1/2}
		\nonumber
		\\
		&&\times\left(\frac{64}{A}\right)^{1/2}
		\left(\frac{10\textrm{ km}}{R}\right)
		\left[\frac{l(l+1)}{2}\right]^{1/2}.
\end{eqnarray}
For an accreting NS with a relatively cold ocean (similar to BC95 or
what we consider in \S 4.1), this frequency is between the g-mode
frequencies, $\lesssim6\textrm{ Hz}$ (BC95), and the f-mode frequency,
$\sim10^3\textrm{ Hz}$, an otherwise empty part of the frequency
domain. Furthermore, the interface mode frequency is independent
of $Z$ and has a simple scaling with both $A$ and $T$, making this
mode a useful diagnostic for measuring the properties of the NS
crust.

\section{Numerical Solutions}

  We now compute eigenfunctions and frequencies for the interface mode
using numerical integrations of the mode equations on a two-component,
ocean/crust NS model. In \S 4.2 we show that these compare very well
with the analytic work of \S 3.
% We then extend our analysis to a NS envelope with a hot surface component
% in \S 4.3 to understand how the interface mode reacts when there exist g-modes
% of comparable frequency.

\subsection{NS Ocean and Crust}

  We assume that the NS has a mass of $M=1.4M_\odot$ and radius of
$R=10\textrm{ km}$. The two components which make up our NS envelope
are a liquid ocean and a degenerate crystalline crust (see Figure 1).
For accreting NSs, the composition of each of these components is
determined by nuclear burning processes which occur on the surface,
especially the rp-process during type I X-rays bursts
(Wallace \& Woosely 1981; Schatz et al. 1999, 2002)
and the hotter burning during superbursts (Cumming \& Bildsten 2001;
Schatz, Bildsten \& Cumming 2003). Woosley et al. (2004) follow the
accretion and subsequent unstable burning over multiple type I bursts
in numerical simulations. Estimating
their results for an accreted composition of 0.05 solar metallicity and an
accretion rate of $1.75\times10^{-9}M_\odot\textrm{ yr}^{-1}$ (which they
denote model ``zM'') after 14 bursts, we assume a composition with mass
fractions of $^{64}$Zn(0.56), $^{68}$Ge(0.35), $^{104}$Pd(0.07)
and $^{106}$Cd(0.02), for both the ocean and crust. At sufficient
pressure and density, this material will change from liquid to solid. This
transition occurs at $\Gamma\approx173$ (see \S 2.1) which implies
a critical density of
\begin{eqnarray}
        \rho_{\rm c} &=& 2.5\times10^9\textrm{ g cm}^{-3}
		\left(\frac{T_8}{4}\right)^3
		\nonumber
		\\
		&&\times
		\left(\frac{A}{64}\right)
		\left(\frac{30}{Z}\right)^6
		\left(\frac{\Gamma}{173}\right)^3,
\end{eqnarray}
for an OCP. We assume that this is a good enough approximation even
though we use a multi-component composition. Due to the strong dependence
of this critical density on $Z$, in cases where burning results in ashes
of especially heavy elements (such as the rp-process at high accretion rates
or during superbursts) crystallization can happen at significantly
shallower depths. This decreases the energy needed for
large mode amplitudes.

  The envelope profile is described by the equation of radiative
transfer, which after substituting $y$ is
\begin{eqnarray}
	F=\frac{4acT^3}{3\kappa}\frac{dT}{dy},
\end{eqnarray}
where $a$ is the radiation constant and $\kappa$ is the opacity. The
opacity is set using electron-scattering (Paczy\'{n}ski 1983),
free-free (Clayton 1993 with the Gaunt factor of Schatz et al. 1999),
and conductive opacities (Schatz et al. 1999 using the basic form of
Yakovlev \& Urpin 1980) in the liquid region. In the crust we make
no distinction and use the same conductive opacity.

  Equation (39) is integrated starting at an outer column depth
($y\approx10^7\textrm{ g cm}^{-2}$) and into
the envelope. The outer temperature is set near $10^8\textrm{ K}$, a
value for which the solutions are minimally sensitive because
of their radiative zero nature (Schwarzschild 1958, or see the discussion
in Piro \& Bildsten 2004). Previous studies of constantly accreting NSs
have shown that the interior thermal balance is set by electron captures,
neutron emissions, and pycnonuclear reaction in the inner crust
(Miralda-Escud\'{e}, Paczy\'{n}ski \& Haensel 1990; Bildsten \& Brown 1997;
Brown \& Bildsten 1998) which release
$\approx1\textrm{ MeV}/m_p\approx10^{18}\textrm{ erg g}^{-1}$
(Haensel \& Zudunik 1990). Depending on the accretion rate and thermal
structure of the crust, this energy will either be conducted into the
core or released into the ocean such that for an Eddington accretion rate
up to $\approx92\%$ of the energy is lost to the core
and exists as neutrinos (Brown 2000).
We therefore set the flux in the ocean to a fiducial value of
$10^{21}\textrm{ erg s}^{-1}\textrm{cm}^{-2}$
(as expected for an accretion rate of $\sim10^{-9}M_\odot\textrm{ yr}^{-1}$,
about one-tenth the Eddington rate).
We solve for $\rho$ using the analytic equation of
state from Paczy\'{n}ski (1983) which is applicable to both the ocean
and crust. The profiles are continuous across the ocean/crust
interface because we assume the same conductive opacity in
each region. Figure 2
shows the envelope profile for the variables most important for
determining the mode properties. The dashed line denotes the
location of the crust (which we assume
is at a column depth of $y_{\rm c}=10^{13}\textrm{ g cm}^{-2}$,
near a density of $\rho_{\rm c}\approx10^{9}\textrm{ g cm}^{-3}$).
\begin{figure}
\epsscale{1.15}
\plotone{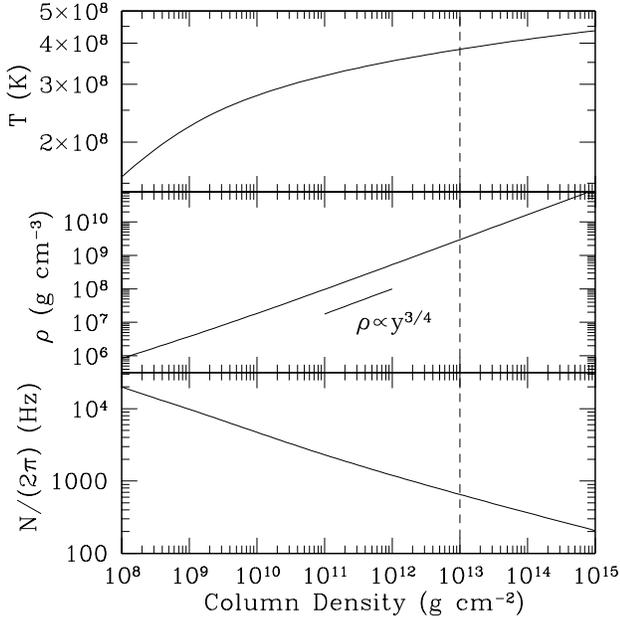}
\figcaption{Temperature, density, and Brunt-V\"{a}is\"{a}l\"{a} frequency
for our two-component neutron star envelope. The dashed line denotes
the depth of the ocean/crust boundary, which lacks a
discontinuity because we use the same conductive opacity in
each region. The flux is assumed to be
constant at $F=10^{21}\textrm{ erg s}^{-1}\textrm{ cm}^{-2}$ with no
other heat sources. This is a fiducial value expected for accreting
neutron stars which have their internal thermal balance set by
electron captures, neutron emissions, and pycnonuclear reactions in the
inner crust. These reactions release $\approx1\textrm{ MeV}/m_p$
with typically $\approx0.1\textrm{ MeV}/m_p$ making its way into the
upper crust (Brown 2000). The power law of the density
($\rho\propto y^{3/4}$) shows that the pressure in the crust is set
by degenerate, relativistic electrons.}
\end{figure}

\subsection{Numerical Integrations and Comparisons with Analytics}

  Using the envelope profile of \S 4.1, we numerically solve
for the interface wave eigenfunctions and eigenfrequencies. We treat each
integration as a two eigenvalue shooting problem, first choosing
$\omega$ and $\lambda$ and then shooting for the bottom boundary
condition. We integrate equations (12) and (13), beginning at
the top of the ocean at a column of $y_{\rm t}\approx10^8\textrm{ g cm}^{-2}$
(where the unstable burning of X-ray bursts typically occurs; Bildsten 1998)
where we set $\Delta P=0$. The boundary conditions at the ocean/crust
interface are set by requiring $\xi_z$ to be continuous,
$\Delta\sigma_{xz}=0$ (which sets $d\xi_x/dz$), and $\Delta\sigma_{zz}$
continuous across the boundary (which sets $d\xi_z/dz$). The discontinuity
of $\xi_x$ is set using $\lambda$. We then integrate equations (15) and (16)
into the NS crust (we assume that $d\mu_0/dz\approx d\Gamma_1/dz\approx0$
and simply set $\mu_0=10^{-2}$
for these examples) down to a depth within the crust where we require
both $\xi_x$ and $\xi_z$ to be zero. We leave the bottom boundary depth
as a free parameter and study how the eigenvalues and eigenfunctions
change with this variable.

  In Figure 3 we show example eigenfunctions with a bottom boundary at
$y_{\rm b}=10^{16}\textrm{ g cm}^{-2}$.
On the same plot we also compare
the analytic eigenfunctions from \S 3.2 and Appendix B. These use the
numeric envelope model to set the local
scale height and $\Gamma_1=1.338$ (its value at the ocean/crust boundary)
to set the eigenvalues. The numerical eigenvalues are
$\omega/(2\pi)=23.43.\textrm{ Hz}$ and $\lambda=-1.5\times10^{-4}$,
which are close to the analytic values of
$\omega/(2\pi)=24.46\textrm{ Hz}$ and $\lambda=-1.8\times10^{-4}$.
As shown in
Figure 1, the radial displacement is found to be very large at the
interface, relative to its value at the ocean surface.
The energy density
(per logarithm of pressure) is
\begin{eqnarray}
        \frac{dE}{d\log y}
        = \frac{1}{2}4\pi R^2 \omega^2\xi^2 y,
\end{eqnarray}
where $\xi^2=|\xi_x|^2+|\xi_z|^2$,
which shows where the mode ``lives.''
The energy density is also strongly concentrated at the interface,
consistent with the result that the mode's properties are set by the
envelope at this depth.
\begin{figure}
\epsscale{1.2}
\plotone{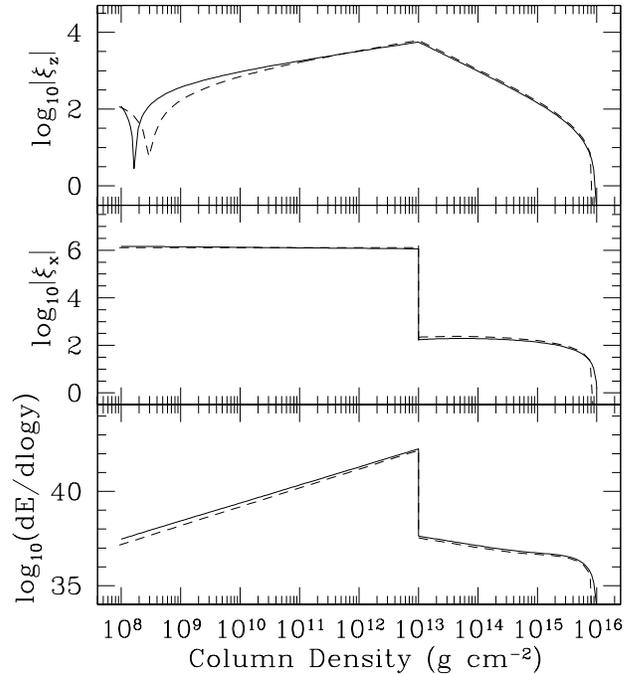}
\figcaption{A comparison of the numerical eigenfunctions (solid curves)
and the analytic approximations (dashed lines) for the interface mode
using the envelope shown in Figure 2, with the bottom boundary set
at $y_{\rm b} = 10^{16}\textrm{ g cm}^{-2}$. 
The numerical solutions find
$\omega/(2\pi)=23.43\textrm{ Hz}$ and $\lambda=-1.5\times10^{-4}$,
while the analytic values are
$\omega/(2\pi)=24.46\textrm{ Hz}$ and $\lambda=-1.8\times10^{-4}$,
using equations (21) and (B7), respectively.
The analytic solutions use the
numeric envelope to set the local scale height, and $\Gamma_1=1.369$
to set the eigenvalues (this is the value of $\Gamma_1$ at the base
of the ocean). The radial eigenfunction, $\xi_z$, shows a single node
in the ocean and a large displacement at the
crust ($y_{\rm c}=10^{13}\textrm{ g cm}^{-2}$), consistent with the discussion
in \S 1. The energy density (bottom panel) shows that the mode energy is
concentrated in the ocean.}
\end{figure}

  As a comparison, we plot $\xi_z$ for both the correct and the ``solid''
ocean/crustal boundary conditions in Figure 4. For the latter we
find a frequency of $\omega_0/(2\pi)=179.62\textrm{ Hz}$. The linear plot
of the amplitudes shows how extreme the radial displacement is at the
crust for the interface mode. We omit comparing $\xi_x$ and the energy density
for the modes since they are almost identical in the ocean.
\begin{figure}
\epsscale{1.2}
\plotone{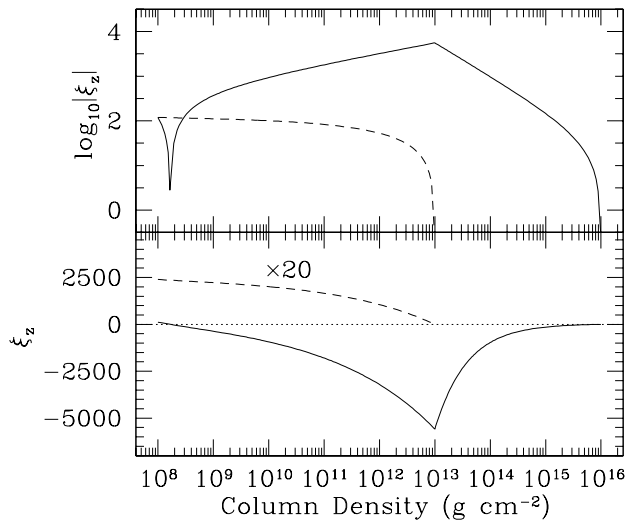}
\figcaption{A comparison of the $\xi_z$ eigenfunction between
a correctly calculated interface wave (solid curve) and the case
of when the crustal boundary is set to $\xi_z=0$ (a standard shallow
ocean wave; dashed curve).
The latter case has an eigenfrequency of $\omega_0/(2\pi)=179.62\textrm{ Hz}$
and no node in the ocean, in contrast to the interface wave's
eigenfrequency of $\omega/(2\pi)=23.43\textrm{ Hz}$.
The linear plot of the displacements (bottom panel) emphasizes the large
displacement at the boundary for the interface wave (note
the other mode is magnified by a factor of 20).}
\end{figure}

  To study the dependence of the mode's properties on
the bottom boundary condition within the crust, we compare
the numeric and analytic
results as a function of the bottom boundary depth, $y_{\rm b}$, in
Figure 5. On the left hand side is the limit when there is no
penetration into the crust, in
this case both the ratio $\xi_{z,\rm c}/\xi_{z,\rm t}$ and $\lambda$
are zero and $\omega/(2\pi)\sim200\textrm{ Hz}$, the values which
would be approximated for a shallow surface wave
(see Figure 4). As the bottom boundary is set deeper, the frequency
decreases, becoming nearly, but not quite equal to
$\sim\mu_0^{1/2}\times200\textrm{ Hz}\sim20\textrm{ Hz}$.
\begin{figure}
\epsscale{1.15}
\plotone{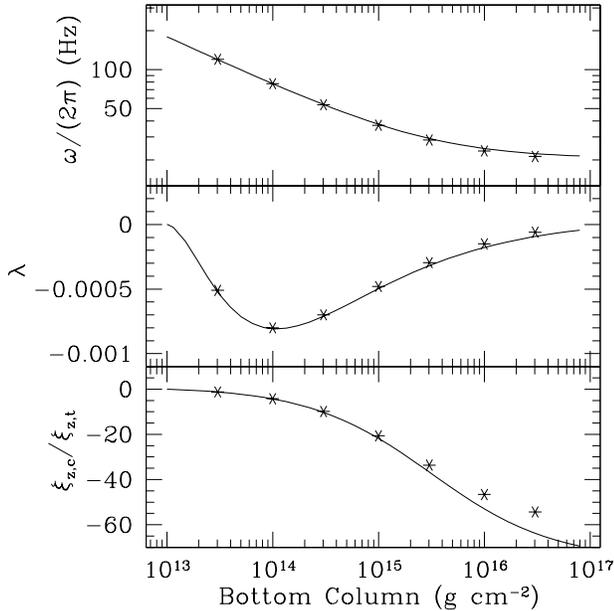}
\figcaption{The properties of the interface mode as a function of the
bottom boundary depth, using the envelope of Figure 2 with the crust at
$y_{\rm c}=10^{13}\textrm{ g cm}^{-2}$. The solid curves show the analytic
estimates, using equations (21), (B7) and (33) (from top to bottom),
while the points
mark the numerical calculations. On the far left we find what is expected
for a shallow ocean wave, and as the curves move right (to larger depths)
they become more like an interface wave.}
\end{figure}

\section{Discussion and Conclusions}

  We have discussed the ocean/crust interface wave, investigating
its properties both analytically and numerically for a two-component
NS envelope. We show that it can be understood as a shallow surface
wave, with a different radial displacement boundary condition at
the bottom of the ocean. This reduces the frequency by roughly
$\sim\mu_0^{1/2}\sim0.1$ from what would be expected if the crust
were instead ``solid.'' In general the interface wave is a global mode
of oscillation, so that the entire star must be taken into account
to calculate the mode to high precision (see discussion in Appendix C).
Fortunately, by using our
analysis, a fairly good estimate of the mode frequency can be made
just from considering the NS properties right near the ocean/crust
interface because the majority of mode energy ($\gtrsim99\%$) resides
in the NS ocean.

  The interface wave should be included in future studies of nonradial
oscillations on both accreting and bursting NSs. For a non-rotating NS
this would result in frequencies of
$\sim10-20\textrm{ Hz}$ which would
then be altered and split due to the NS spin
(Bildsten, Ushomirsky \& Cutler 1996). The observation of multiple mode
periods would greatly assist our understanding of the NS crust.
The g-modes are sensitive to the depth of the crust (since it provides a
bottom boundary), while the interface wave is sensitive to both the depth
and the value of the shear modulus. This provides information about
the temperature and
composition of the crust, which is determined by unstable nuclear burning
processes (X-ray bursts and superbursts) that happen in the ocean.

  Since the example considered in this work
is a relatively ``cool'' envelope, the g-mode frequencies
($\lesssim6\textrm{ Hz}$) are well separated from the interface wave
frequency ($\sim10-20\textrm{ Hz}$). On a bursting NS, which has
temperatures in the upper atmosphere as high as
$10^9\textrm{ K}$, at least one g-mode has a frequency higher than the
interface wave. The interface wave would then have an additional node
in its eigenfunction,
due to the relation between g-mode frequency ordering and number of nodes,
but its frequency would be unchanged as long as the ocean
stays at a temperature of $\sim4\times10^8\textrm{ K}$ (since this is where the
mode's energy is concentrated). Woosley et al. (2004) show that the thermal
wave during an X-ray burst does not penetrate deeply into the
ocean, so that our estimates for the interface mode frequency still hold.

  The other important astrophysical context where our work can be applied
is for old and/or massive WDs that have crystalline cores. This may be
seen in the pulsations of BPM 37093 (Kanaan et al. 1992), a high-mass
ZZ Ceti star that should be considerably crystallized (Winget et al. 1997).
There is also the exciting possibility that some of the pulsators
discovered through the Sloan Digital Sky Survey (Mukadam et al. 2004) may
be massive enough to be crystallized. Scaling equation (37) to
the fiducial interior properties of a $\approx1.1M_\odot$ WD,
we find a crustal mode period of
\begin{eqnarray}
        P&\approx&95\textrm{ s}
                \left(\frac{173}{\Gamma}\right)^{1/2}
                \left(\frac{10^7\textrm{ K}}{T}\right)^{1/2}
                \nonumber
                \\
                &&\times\left(\frac{A}{16}\right)^{1/2}
                \left(\frac{R}{5\times10^8\textrm{ cm}}\right)
                \left[\frac{2}{l(l+1)}\right]^{1/2},
\end{eqnarray}
which for the $l=1$ mode is between the periods of the f-mode
($\sim10-20\textrm{ s}$) and g-modes ($\gtrsim300\textrm{ s}$).
HV79 (and more recently Montgomery \& Winget 1999) previously calculated
the spectrum of g-modes and torsional modes expected for these stars,
concentrating on the effects of crystallization. HV79 do not single out any
one mode as a ``crustal mode,'' but in their crystallized models their lowest
order g-mode (which they denote $_{-1}S_{l}$) matches many of the properties
we identify as being characteristic of a crustal mode. This includes a large
radial displacement at the crust and a nearly constant transverse displacement
in the WD ``ocean'' (see Figures 1 and 2 of HV79).
% Depending on the depth of crystallization, the crustal mode's frequency
% may appear near the other g-modes (as discussed directly above for NS
% X-ray bursts) or perhaps well-separated,
% between the g-modes and f-mode (as for the NS model from \S 4).
Furthermore, HV79 find the period of their
$_{-1}S_{l}$ mode to be between the f-mode and the other g-mode periods, in
confirmation of our above discussion.

  We thank both Phil Arras and Omer Blaes for helpful discussions.
We also thank the referee, Michael Montgomery, for pointing out the 
HV79 paper and its relevance to our work.
This work was supported by the National Science Foundation under
grants PHY99-07949 and AST02-05956, and by the Joint Institute
for Nuclear Astrophysics through NSF grant PHY02-16783.

\begin{appendix}

\section{Influence of a Finite Shear Modulus}

  The interface wave's properties are set by how it is influenced by the
finite shear modulus of the NS crust. Unlike g-modes, which see the
crust as solid, the interface wave reacts to the crust as if it is
a flexible and compressible surface. We now explain why each
of these cases occur.

  Using $\mu_0\equiv\mu/P$,
and assuming that $d\mu_0/dz\approx d\Gamma_1/dz\approx0$,
equations (15) and (16) can be rewritten as
\begin{eqnarray}
        -\frac{d^2\xi_z}{dz^2} \left( \Gamma_1 + \frac{4}{3}\mu_0\right)
        &=&\xi_z \left( \frac{\omega^2}{gh} - k_x^2\mu_0 \right)
                - \frac{d\xi_z}{dz} \left( \frac{4}{3}\frac{\mu_0}{h}
                +\frac{\Gamma_1}{h} \right)
        \nonumber
        \\
        &&+ik_x\xi_x \left[ \frac{(1-\Gamma_1)}{h}
                + \frac{2}{3}\frac{\mu_0}{h}\right]
        \nonumber
        \\
        &&+ ik_x\frac{d\xi_x}{dz}\left( \frac{\mu_0}{3} + \Gamma_1 \right),
\end{eqnarray}
and
\begin{eqnarray}
        -\mu_0 \frac{d^2\xi_x}{dz^2}
        &=&\xi_x \left( \frac{\omega^2}{gh} - \frac{4}{3}k_x^2\mu_0
                - \Gamma_1k_x^2\right)
                - \frac{\mu_0}{h}\frac{d\xi_x}{dz}
        \nonumber
        \\
        &&-ik_x\xi_z \left( \frac{\mu_0}{h}+\frac{1}{h} \right)
                +ik_x\frac{d\xi_z}{dz} \left( \frac{\mu_0}{3}+\Gamma_1 \right).
\end{eqnarray}
The simplest analysis of these equations arises when
perturbations scale like $\sim \exp(ik_z z)$ and dispersion relations
are in the WKB limit (when the radial wavenumber
$k_z\gg1/h$). This results in toroidal and spheroidal modes with
frequencies
\begin{eqnarray}
        \omega^2_{\rm shear} &=& \mu_0(P/\rho)k^2,
        \\
        \omega^2_{\rm sound} &=& (\Gamma_1+4\mu_0/3)(P/\rho)k^2,
\end{eqnarray}
where $k^2=k^2_x+k^2_z$ is the total wavenumber. The first
is a shear mode (with a frequency of $\sim10^2-10^3\textrm{ Hz}$)
and the second is a sound wave slightly modified
by $4\mu_0/3$ ($\sim10^4-10^5\textrm{ Hz}$).
These imply that a g-mode with a frequency of $\lesssim6\textrm{ Hz}$
(BC95) will be reflected at the interface because of the frequency mismatch.
Unfortunately, this analysis cannot be extended
to the interface mode because it has $k_z\lesssim1/h$, in contradiction
to the WKB limit.

  To see that a mode with a constant transverse velocity (like the interface
wave) is not excluded from the crust we compare the transverse
momentum terms from  equation (9). The shear modulus term has two pieces, the
larger one is $d\delta\sigma_{xz}/dz\sim\mu d^2\xi_x/dz^2$.
This term strongly affects the mode's eigenfunctions
when it is comparable to the transverse acceleration of the mode,
$\rho\omega^2\xi_x$. Using $d/dz\sim k_z$ this implies a critical shear
modulus of
$\mu_{\rm crit}/P\sim \omega^2/(ghk_z^2)$. In the case
of g-modes $k_z\sim1/h$ and
$\omega^2\approx4N^2(hk_x)^2$ so that the critical modulus is (BC95)
\begin{eqnarray}
	\frac{\mu_{\rm crit}}{P}(\textrm{g-mode})
	\sim \frac{6l(l+1)}{Z}\frac{k_{\rm B}T}{E_{\rm F}}
	\left( \frac{h}{R} \right)^2.
\end{eqnarray}
Substituting values near the ocean/crust interface for degenerate,
relativistic electrons,
\begin{eqnarray}
        \frac{\mu_{\rm crit}}{P}(\textrm{g-mode})
        \sim 10^{-11}T_8^2l(l+1)
	\left(\frac{64}{A}\right)^2
	\left(\frac{30}{Z}\right)^{2/3}
        \left( \frac{\Gamma}{173} \right),
\end{eqnarray}
which is clearly less than the value of $\mu/P$ from equation (6), implying
that the g-modes will be excluded from the crust.

  One the other hand, the interface wave has a constant transverse velocity
so that $k_z\sim1/R$. Using $\omega^2\approx\mu_0ghk^2_x$, the interface
wave has a critical shear modulus of
\begin{eqnarray}
        \frac{\mu_{\rm crit}}{P}(\textrm{interface wave})
	\sim \mu_0 l(l+1),
\end{eqnarray}
which is just the actual shear modulus present at the boundary,
so the wave must penetrate the crust. This shows that in the interface wave
the transverse momentum is perfectly balanced by the shear stress,
so that the ocean/crust boundary behaves like a flexible and
compressible surface.

\section{Transverse Crustal Eigenfunction and Discontinuity Eigenvalue}

  To estimate the transverse crustal eigenfunction, we order the terms
from equation (A2) and drop those which are small.  Using equations (24)
and (26) we first relate the ordering of the transverse and radial
displacements within the curst.
Dividing equation (A2) through by $\xi_z$, multiplying through by $h^2$
and using the above relations, the ordering of each of the terms are
\begin{eqnarray}
        \mathcal{O}( \mu_0 k_x h )
                &\sim&\mathcal{O}( \omega^2h^2k_x/g )
                        +\mathcal{O}( \mu_0 k_x^3h^3 )
                        +\mathcal{O}( k_x^3h^3 )
                        +\mathcal{O}( k_xh )
        \nonumber
        \\
                &&+\mathcal{O}( \mu_0k_xh )
                        +\mathcal{O}( k_xh )
                        +\mathcal{O}( \mu_0k_xh )
                        +\mathcal{O}( k_xh ).
\end{eqnarray}
The size of these quantities are approximated using $k_xh\sim10^{-4}-10^{-3}$ and
$\mu_0\approx10^{-2}$. Just as with the radial equation
considered in \S 3.2, the terms which contain $\omega^2$ are negligible
(as is seen from considering $\omega^2$ anywhere in the range of
$\sim\mu_0ghk_x^2-ghk_x^2$). To find a relation which includes $\xi_x$ we must
consider both first- and second-order terms of equation (A2) to get
\begin{eqnarray}
        -\mu_0 \frac{d^2\xi_x}{dz^2}
                &=&- \frac{\mu_0}{h}\frac{d\xi_x}{dz}
                -ik\xi_z \left( \frac{\mu_0}{h}+\frac{1}{h} \right)
                \nonumber
                \\
                &&+ik\frac{d\xi_z}{dz} \left( \frac{\mu_0}{3}+\Gamma_1 \right),
\end{eqnarray}
as a new simplified differential equation.

  We substitute the eigenfunction for $\xi_z$, equation (28)
into equation (B2) to solve for
$\xi_x$. The two constants of integration are set by imposing that
$\Delta\sigma_{xz}=0$ at the top of the crust and that $\xi_x$
(just like $\xi_z$) goes to zero at $s_{\rm b}$, so that
\begin{eqnarray}
        \xi_x(s) = \frac{ik_xs_{\rm b}\xi_{z,\rm c}}{(s_{\rm b}/s_{\rm c})^3-1}
                \left\{ \alpha\left[\left(\frac{s_{\rm b}}{s}\right)^3-1\right]
                        +\beta\left[\left(\frac{s_{\rm b}}{s}\right)^2-1\right]\right.
                \nonumber
                \\
                \left.+\gamma\left[1-\frac{s}{s_{\rm b}}\right] \right\},\hspace{0.3cm}
\end{eqnarray}
where we have introduced dimensionless constants that only depend
on the star,
\begin{eqnarray}
        \alpha
        &\equiv& \frac{2\mu_0+4-3\Gamma_1}{3\mu_0}\frac{s_{\rm c}}{s_{\rm b}}
        - \frac{1}{3\mu_0}\left( \frac{s_{\rm c}}{s_{\rm b}}\right)^4,
        \\
        \beta &\equiv& -\frac{3\mu_0+4-3\Gamma_1}{2\mu_0},
        \\
        \gamma &\equiv& 1+\frac{1}{\mu_0}.
\end{eqnarray}
Substituting $s=s_{\rm c}$ and dividing by equation (17) we find
\begin{eqnarray}
        \lambda&\equiv&\frac{\xi_{x,\rm t}}{\xi_{x,\rm c}}
        \nonumber
        \\
        &=& \frac{\omega^2s_{\rm b}/g}{(s_{\rm b}/s_{\rm c})^3-1}
                \frac{\xi_{z,\rm c}}{\xi_{z,\rm t}}
          \left\{ \alpha\left[\left(\frac{s_{\rm b}}{s_{\rm c}}\right)^3-1\right]
                        +\beta\left[\left(\frac{s_{\rm b}}{s_{\rm c}}\right)^2-1\right]\right.
                \nonumber
                \\
                &&\hspace{4.7cm}\left.+\gamma\left[1-\frac{s_{\rm c}}{s_{\rm b}}\right] \right\},
\end{eqnarray}
for the discontinuity eigenvalue.

  We use equation (B3) (or consider the infinite crust limit of Appendix C),
to see that the size of
$\xi_{x,\rm c}/\xi_{z,\rm c}\sim k_xs_{\rm c}$. This shows that when
we consider terms in equation (30), the $\xi_x$ terms are of order
\begin{eqnarray}
	\frac{k_x\xi_x}{d\xi_z/dz}
	\sim (k_x h_{\rm c})^2\ll 1,
\end{eqnarray}
smaller than the $\xi_z$ terms and can thus be ignored.

\section{Infinite Crust Limit}

  Just as we did for the radial eigenfunction for equation (29),
we take the $s_{\rm b}\gg s_{\rm c}$ limit on equation (B3)
to find the transverse eigenfunctions in the infinite crust limit,
\begin{eqnarray}
        \xi_x(s) &=& ik_xs_{\rm c}\xi_{z,\rm c}
                \left[\frac{2\mu_0+4-3\Gamma_1}{3\mu_0}
                \left( \frac{s_{\rm c}}{s}\right)^3\right.\hspace{0.7cm}
                \nonumber
                \\
                &&\hspace{1.5cm}\left.-\frac{3\mu_0+4-3\Gamma_1}{2\mu_0}
                \left( \frac{s_{\rm c}}{s}\right)^2\right],
                \nonumber
                \\
                &\approx& ik_xs_{\rm c}\xi_{z,\rm c}
                \left[\frac{2}{3}
                \left( \frac{s_{\rm c}}{s}\right)^3
                -\frac{3}{2}
                \left( \frac{s_{\rm c}}{s}\right)^2\right],
\end{eqnarray}
which corresponds to
\begin{eqnarray}
        \lambda &=& -\frac{\omega^2s_{\rm c}}{g}
                \frac{\xi_{z,\rm c}}{\xi_{z,\rm t}}
                \left(\frac{5\mu_0+4-3\Gamma_1}{6\mu_0}\right),
                \nonumber
                \\
                 &\approx&-\frac{5}{6}\frac{\omega^2s_{\rm c}}{g}
                \frac{\xi_{z,\rm c}}{\xi_{z,\rm t}},
\end{eqnarray}
for the discontinuity eigenvalue.

  We use these eigenfunctions to estimate the mode's energy density,
$dE/d\log y$, from equation (37). We consider the limit of
$s\rightarrow\infty$ to see if the energy density falls off with increasing
depth. If this happens, it implies a natural depth at which to set a
lower boundary and provides confidence that the interface mode
is determined by conditions local to the ocean/crust boundary.
Comparing equations (29) and (C1) we find that at depths of order
the NS radius $\xi_x$ will dominate so that $\xi^2\propto1/s^4$.
For fully degenerate and relativistic material
$P\propto y \propto \rho^{4/3}$.
Using $h=P/(\rho g)$ and $h\propto s$ we find that $y\propto s^4$.
This means that at the greatest possible depths
$dE/d\log y\sim\xi^2y\sim\textrm{constant}$, so that, unfortunately,
deep regions of the NS much be taken into account to calculate the
interface mode to very high accuracy. In practice, this is not necessary
because there are structural changes within a realistic NS crust
that introduce new bottom boundaries. Due to the large amount of
energy concentrated in the ocean ($\gtrsim99\%$, see Figure 3),
our frequency estimates are still good even with this difficulty.

\end{appendix}

\end{document}